
\input harvmac


\lref\MatoneRR{
M.~Matone and L.~Mazzucato,
arXiv:hep-th/0307130.
}

\lref\MatoneUY{
M.~Matone,
math.ag/0105051.
}

\lref\BonoraWN{
L.~Bonora and M.~Matone,
Nucl.\ Phys.\ B {\bf 327}, 415 (1989).
}

\lref\MatoneTJ{
M.~Matone,
Int.\ J.\ Mod.\ Phys.\ A {\bf 10}, 289 (1995)
hep-th/9306150.
}

\lref\KrausJF{ P.~Kraus and M.~Shigemori,
JHEP {\bf 0304}, 052 (2003), hep-th/0303104.
}
\lref\AldayGB{
L.~F.~Alday and M.~Cirafici,
JHEP {\bf 0305}, 041 (2003), hep-th/0304119.

P.~Kraus, A.~V.~Ryzhov and M.~Shigemori,
JHEP {\bf 0305}, 059 (2003), hep-th/0304138.
}

\lref\AganagicXQ{
M.~Aganagic, K.~Intriligator, C.~Vafa and N.~P.~Warner,
hep-th/0304271.
}
\lref\CachazoKX{ F.~Cachazo,
hep-th/0307063.
}

\lref\MatoneRR{
M.~Matone and L.~Mazzucato,
JHEP {\bf 0310}, 011 (2003), arXiv:hep-th/0307130.
}

\lref\CDSW{ F.~Cachazo, M.~R.~Douglas, N.~Seiberg and E.~Witten,
JHEP {\bf 0212}, 071 (2002), hep-th/0211170.
}

\lref\WittenYE{ E.~Witten,
hep-th/0302194.
}

\Title{\vbox{\baselineskip11pt\hbox{hep-th/0307285}
\hbox{DFPD/03/TH/29} }} {\vbox{ \centerline{The Affine Connection
of Supersymmetric}
\smallskip
\centerline{$SO(N)/Sp(N)$ Theories}
 \vskip
1pt }}
\smallskip
\centerline{Marco Matone}
\smallskip
\bigskip
\centerline{Dipartimento di Fisica ``G. Galilei'', Istituto
Nazionale di Fisica Nucleare,} \centerline{Universit\`a di Padova,
Via Marzolo, 8 -- 35131 Padova, Italy}
\medskip

\bigskip
\vskip 1cm
 \noindent
We study the covariance properties of the equations satisfied by the generating functions
of the chiral operators $R$ and $T$ of
supersymmetric $SO(N)/Sp(N)$ theories with symmetric/antisymmetric tensors. It turns
out that $T$ is an affine connection. As such it cannot be integrated
along cycles on Riemann surfaces. This explains
the discrepancies observed by Kraus and Shigemori.
Furthermore, by means of the polynomial defining the Riemann surface, seen as
quadratic--differential,
one can construct an affine connection that added to $T$ leads to a new generating
function which can be consistently integrated.
Remarkably, thanks to an identity, the original equations are equivalent to equations involving
only one--differentials. This provides a geometrical explanation of the map recently derived by
Cachazo in the case of $Sp(N)$ with antisymmetric tensor.
Finally, we suggest a relation between the
Riemann surfaces with rational periods which arise in studying the Laplacian
on special Riemann surfaces and the integrality condition
for the periods of $T$.

\vskip 0.5cm

\Date{July 2003}
%
\baselineskip14pt

%
%


In this note we consider the problem
of the contributions to the free energy at order $h$, with $h$
the dual Coxeter number, which arises in the case of supersymmetric $SO(N)/Sp(N)$ theories with
symmetric/antisymmetric tensors \KrausJF\AldayGB\AganagicXQ. Recently,
in \CachazoKX, Cachazo derived a map from $Sp(N)$ to a $U(N+2n)$
theory with one adjoint and a degree $n+1$ tree level
superpotential. In his construction he observed interesting geometrical
questions, such as the unusual phenomenon that vanishing of the period of
$T$ through a given cut does not imply that the cut closes on--shell. There
are also other unusual properties.

Here we show that the chiral operator $T$ of supersymmetric
$SO(N)/Sp(N)$ theories with symmetric/antisymmetric tensors is an
affine connection and so it cannot be consistently integrated
along cycles on the Riemann surface. This is the geometrical
origin of the discrepancies observed by Kraus and Shigemori. In
our construction we will see that covariance forces us to consider
the polynomial defining the Riemann surface as a
quadratic--differential. By means of such a differential we
construct an affine connection that added to $T$ leads to a new
generating function $\tilde T$ which is now a true
one--differential and can be consistently integrated. We then show
that, thanks to a remarkable identity, the original equations are
equivalent to equations involving only one--differentials. We will
conclude by suggesting a relation between the special Riemann
surfaces with rational periods introduced in \MatoneUY\ and the
integrality condition for the periods of $T$.

Let us start by considering the ${\cal N}=1$ supersymmetric theory
with gauge group $Sp(N)$ with a chiral superfield in the
antisymmetric representation. The classical superpotential is
\eqn\wtree{ W_{tree}=\sum_{j=0}^n{g_j\over j+1} \Tr\,\Phi^{j+1}. }
The generating functions of chiral operators \eqn\generatori{
T(z)=\Tr\, {1\over z-\Phi},\qquad R(z)=-{1\over32\pi^2}\Tr\,
{W_\alpha W^\alpha\over z-\Phi}, } satisfy the equations
\eqn\leequazioni{ [W'R]_-={1\over2}R^2,\qquad [W'T]_-=TR+2R'. }
Usually, both $T$ and $R$ are integrated over cycles on the
Riemann surface. It follows that both $R$ and $T$ should be
one--differentials. Let us show that this is not the case. More
precisely, we show that the unique consistent geometric
interpretation of \leequazioni\ is that $R$ be a one--differential
and $T$ an affine connection. Actually, note that the unique
interpretation consistent with the first equation in \leequazioni\
is that $R$ be a one--differential. Let us now consider the right
hand side of the second equation in \leequazioni. This contains
the first derivative of the one--differential $R$. In the
intersection between two patches we have \eqn\intersezione{
R_\alpha(z_\alpha)={dz_\beta\over dz_\alpha} R_\beta(z_\beta), }
so that the transformation of $R'$ is \eqn\trasf{ {d\over
dz_\alpha}R_\alpha(z_\alpha)=\Bigl({dz_\beta\over
dz_\alpha}\Bigr)^2 {d\over dz_\beta}R_\beta(z_\beta)+{d^2
z_\beta\over d z_\alpha^2}R_\beta(z_\beta). } In order that
$[W'T]_-=TR+2R'$ be covariantly defined, we should relax the
condition that $T$ be a one--differential. Actually, we should
require that $TR+2R'$ be a quadratic differential, that is
\eqn\quadratic{ T_\alpha(z_\alpha)R_\alpha(z_\alpha)+2{d\over
dz_\alpha}R_\alpha(z_\alpha)= \Bigl({dz_\beta\over
dz_\alpha}\Bigr)^2\Bigl[ T_\beta(z_\beta)R_\beta(z_\beta)+2{d\over
dz_\beta}R_\beta(z_\beta)\Bigr]. } Eqs.\trasf\ and \quadratic\
give \eqn\ttrasf{T_\beta(z_\beta){dz_\beta\over
dz_\alpha}=T_\alpha(z_\alpha)+ 2{dz_\alpha\over
dz_\beta}{d^2z_\beta\over dz_\alpha^2}. } Recalling that an affine
connection on a Riemann surface $\Sigma$ is a set of functions
${\cal A}$, each one defined on a patch, such that in the
non--empty  intersection $U_\alpha \cap U_\beta$ \eqn\affine{
{\cal A}_\beta(z_\beta){dz_\beta\over dz_\alpha}={\cal
A}_\alpha(z_\alpha) +{d\over dz_\alpha}\ln {dz_\beta\over d
z_\alpha}, } we see that $T$ turns out to be twice an affine
connection. As such it cannot be consistently integrated on cycles
on the Riemann surface. This explains the discrepancies found
between the gauge theory calculations and the ones obtained
by the loop equation formulations. This also
explains why one obtains the geometrical strange effect that the
vanishing of the period of $T(z)dz$ is non zero but the cut closes
up on--shell.

In order to write a covariantized version of Eq.\leequazioni\ we
can use the procedure introduced in \BonoraWN\ in formulating the
KdV equation on a Riemann surface, leading to $W$--algebras, and
further developed in \MatoneTJ\ to introduce higher order cocycles
on Riemann surfaces. Furthermore, it has been shown in \MatoneTJ\
that this covariantization is strictly related to uniformization
and Liouville theories. Let us shortly review it. Let us consider
the negative power of the Poincar\'e metric \eqn\ek{
e^{-k\varphi}=
|{J_H^{-1}}'|^{-2k}\left({J_H^{-1}-\overline{J_H^{-1}}\over 2i}
\right)^{2k}, } where ${J_H^{-1}}$ is the inverse of the
uniformizing map from the upper half plane to $\Sigma$. One can
easily check that the negative powers of the Poincar\'e metric
satisfy the higher order generalization of nullvector equations,
that is  \MatoneTJ\ \eqn\el{ {\cal S}^{(2k+1)}_{J_H^{-1}}\cdot
e^{-k\varphi} =0, \qquad k=0,{1\over 2}, 1,\ldots, } with ${\cal
S}^{(2k+1)}_h$ the higher order covariant Schwarzian operator
\eqn\em{ {\cal S}^{(2k+1)}_h=(2k+1) (h')^k \partial_z
(h')^{-1}\partial_z (h')^{-1}\ldots
\partial_z (h')^{-1}\partial_z (h')^k,
}
where the number of derivatives is $2k+1$.
The univalence of ${J_H^{-1}}$ implies holomorphicity
of the ${\cal S}^{(2k+1)}_{J_H^{-1}}$ operators.
Eq.\el\
is manifestly covariant and singlevalued on
$\Sigma$. Furthermore it can be proved that
the dependence of
${\cal S}^{(2k+1)}_h$ on $h$ appears only through
${\cal S}^{(2)}_h\cdot 1=\{h,z\}$ and its derivatives; for example
\eqn\en{
{\cal S}^{(3)}_{J_H^{-1}}=
3\left(\partial_z^3+2T^F\partial_z+{T^F}'\right),
}
which is the second symplectic structure of the KdV equation, where now
$T^F$ is the Fuchsian projective connection \MatoneTJ.

In the above derivation we used the polymorphic vector fields $1/{J_H^{-1}}'$
to construct the covariant operators ${\cal S}^{(2k+1)}_{J_H^{-1}}$
mapping $(-k,n)$--differentials, to $(k+1,n)$--differentials. In the present
situation we should covariantize the quantity $R'$. However, note that we cannot use
the inverse map of uniformization. The reason is that covariant operators constructed
in terms of ${J_H^{-1}}$ are singlevalued only in the case in which the operator
is of order greater than $2$. So we need a vector field which also encodes the
geometrical information of the underlying Riemann surfaces. To solve this question
we notice that $y^2={W'}^2+f$ contains all the geometrical data. Also, note that
covariance forces us to consider $f$ as quadratic differential and so
$y^2$ itself is a quadratic differential. It follows that the natural candidate to
covariantize the $R'$ is to use the vector field
${1\over y}$. As in the covariantization reviewed above, the covariantized derivative
should be homogeneous in $y$ and map the one--differential $R$ to a quadratic differential.
This unequivocally leads to
\eqn\co{
{\cal D}_z\equiv y{d\over dz}y^{-1}={d\over dz}-{y'\over y},
}
so that
\eqn\erreprimo{
{d\over dz}R\longrightarrow {\cal D}_z R.
}
Let us now rewrite the right hand side of the second equation
in \leequazioni\ in the following form
\eqn\deltat{
TR+2R'=(T+\delta T)R+2 {\cal D}_z R.
}
It follows that
\eqn\dd{
\delta T={d\over dz}\ln y^2.
}
Now note that since both $TR+2R'$ and ${\cal D}_z R$ are quadratic differentials,
and $R$ is a one--differential, it follows that
\eqn\tn{
\tilde T=T+{d\over dz}\ln y^2,
}
is a well--defined one--differential and can be consistently integrated along cycles on the
Riemann surface.

Note that $y=W'-R$ is covariantly constant, that is
the analogue of \el\ is
\eqn\modozero{ {\cal D}_z (W'-R)=0. } It
follows that the second equation in \leequazioni\ can be
equivalently written in the form \eqn\nellaforma{ [W'(\tilde
T-\delta T)]_-=\tilde TR+2 {\cal D}_zW'. } We now show that
\eqn\dellaforma{ [W'\delta T]_-=-2{\cal D}_zW', } so that the
original equation $[W'T]_-=TR+2R'$ is equivalent to
\eqn\perlaforma{ [W'\tilde T]_-=\tilde T R, } which is covariant
and involves only one--differentials. Let us give a closer look to
Eq.\dellaforma. It expresses the fact that the covariant
derivative of $W'$ is obtained by the negative frequencies of
$W'\delta T$. It can be also written in the form \eqn\carina{
[W'{y'\over y}]_-=W'{y'\over y}-W'', } that is \eqn\carine{
W''=[W'{y'\over y}]_+. } On the other hand, differentiating $y^2={W'}^2+f$, we have \eqn\wehave{
W'{y'\over y}=W''+{1\over2}{W'f'-2W''f\over {W'}^2+f}. } The
second term in the right hand side has the form $P_{2n-2}/P_{2n}$, with $P_j$ denoting
a polynomial of order $j$. Expanding in powers of $1/z$, we
see that \eqn\wesee{ \Bigl[{W'f'-2W''f\over {W'}^2+f}\Bigr]_+=0, }
so that \carine\ is satisfied and the original equation is
equivalent to its covariant form Eq.\perlaforma.

Let us now consider the case of gauge group $SO(N)$ with a symmetric tensor. In this case
we have
\eqn\leequazionib{
[W'R]_-={1\over2}R^2,\qquad [W'T]_-=TR-2R'.
}
Repeating the above derivation, we now see that consistency implies that in this case
$T$ transforms
as $-2$ times an affine connection, that is
\eqn\ttrasfson{T_\beta(z_\beta){dz_\beta\over dz_\alpha}=T_\alpha(z_\alpha)-
2{dz_\alpha\over dz_\beta}{d^2z_\beta\over dz_\alpha^2}.
}
It follows that the correct one--differential to integrate is
\eqn\tn{
\tilde T=T-{d\over dz}\ln y^2,
}
whereas the Eqs.\nellaforma\ and \dellaforma\ become
\eqn\nellaformab{
[W'(\tilde T-\delta T)]_-=\tilde TR-2 {\cal D}_zW',
}
and
\eqn\dellaformab{
[W'\delta T]_-=2{\cal D}_zW',
}
which is satisfied because now $\delta T=-{d\over dz}\ln y^2$. It follows that also in this case
we have
\eqn\perlaformab{
[W'\tilde T]_-=\tilde T R.
}

We saw that while classically one has $f=0$, so that $R=0$ and $T$ is a one--form on the Riemann sphere,
quantum mechanically $T$ has anomalous transformations.
This can be modified
to a one--form $\tilde T$ that can be integrated along cycles.
However, since the original $T$ cannot be integrated, one should understand whether
the conditions  $\oint_{A_i}T=N_i$ and $\oint_{B_i}T=b_i$ in
\CachazoKX\ are essential. Actually, it seems that Cachazo's results can be obtained
by imposing the weaker conditions\foot{We follow the notation in \CachazoKX.}
\eqn\weaker{
\Tr\, {\bf 1} = \Tr\, {\bf 1}_U-2n,
}
\eqn\weakertwo{
{1\over 2\pi i}\oint_{A_i}T_U=N_i+2,\qquad i=1,\ldots,n,
}
that do not imply $T$ to be a one--form. So, essentially, the results in \CachazoKX\
do not use the integrality condition on the cycles of the original $T$. This is different
with respect to the previous investigations based on the loop equation \AldayGB. Understanding how the
results in \CachazoKX\ can be obtained in the original formulation of Kraus and Shigemori \KrausJF\ remains
an interesting open question. Even if a detailed and clarifying investigation of the
ambiguities has been given in \AganagicXQ, it seems that there are still interesting
aspects to be fully understood. In this respect it is interesting to observe that besides
the appearance of the affine connection, $SO(N)/Sp(N)$
theories have another distinguished, perhaps related, feature concerning the algebraic
relations satisfied by $S$. In the
case of $SU(N)$  one has \CDSW\
\eqn\identity{S^N=\Lambda^{3N}+\{\overline Q_{\dot\alpha},
X^{\dot\alpha}\}. }
In \KrausJF\ it was suggested that such kind of relations may
be at the origin of the observed discrepancy.
However in \CDSW\ it was pointed out that the identity \identity\ of
the microscopic $SU(N)$ theory corresponds to the field equation
$S^N=\Lambda^{3N}$ one obtains as stationary points of the
superpotential. This dual description is similar to the magnetic description of
free electrodynamics where the Bianchi identity arises only
on--shell. In this respect, as observed by Cachazo \CachazoKX, if
this picture holds, since the effect of the field equation is not
directly visible in the off--shell low energy effective
superpotential, it follows that the identity should not be at the
origin of the discrepancy observed in \KrausJF. However, there is another identity besides
Eq.\identity, that is \MatoneRR\
\eqn\identitytwo{
(S^N-\Lambda^{3N})^N=0.}
It is interesting to find this relation in the case of a generic simple group
$G$. First recall that since $S$ is a bilinear in fermionic fields, classically
we have \CDSW\
\eqn\identitythree{
S^{{\rm dim}(G)+1}=0.
}
On the other hand \CDSW\WittenYE\
\eqn\identityfour{
S^h=\{\overline Q_{\dot\alpha},
X^{\dot\alpha}\},
}
where $h$ is the dual Coxeter number of $G$. This relation gets instantonic corrections
\CDSW\WittenYE\
\eqn\identityfive{
S^h=c(G)\Lambda^{3h}+\{\overline Q_{\dot\alpha},
X^{\dot\alpha}\}.
}
Now note that the $X^{\dot\alpha}$ in \identityfour\ and
\identityfive\ can differ only by a chiral operator: dimensional analysis
and $R$--symmetry forbid the existence of terms $\{\overline
Q_{\dot\alpha},\delta X^{\dot\alpha}\}$ that vanish as
$\Lambda\to0$. However, while how observed in  \MatoneRR\ in the case of
$SU(N)$ the last two identities imply $\{\overline Q_{\dot\alpha},
X^{\dot\alpha}\}^N=0$, that is Eq.\identitytwo,
for other groups there is an interesting new phenomenon. Namely,
for general groups we have
\eqn\identitysix{
\{\overline Q_{\dot\alpha},
X^{\dot\alpha}\}^M=0,\qquad
}
where $M$ is the lowest integer such that $hM\geq {\rm dim}(G)+1$. Eq.\identitysix\
implies
\eqn\identityseven{
(S^h-c(G)\Lambda^{3h})^M=0,}
that in the classical limit reduces to
\eqn\identityeight{
S^{hM}=0.
}
Of course, since $S^{{\rm dim}(G)+1}=0$ implies $S^{hM}=0$ even in the case
in which $hM>{{\rm dim}(G)+1}$, it follows that Eqs.\identitythree\
and \identityeight\ are consistent. Nevertheless, it is interesting to observe that
in general, as in the case of $Sp(N)$ where $h=N/2+1$, one has\foot{Since
$N(N+1)/2+1=h(N+3)+4$, we see that only for $Sp(10)$ one has $hM={{\rm dim}(G)+1}$.
This special case is the first of the ones not explicitly investigated
in \KrausJF\ and \CachazoKX. It is interesting to observe that since for
$h(SO(N))=N-2$, so that $N(N+1)/2+1=h(N+3)/2+4$, we have that also in this case
$N=10$ is the unique solution of $hM={{\rm dim}(G)+1}$.}
$hM>{{\rm dim}(G)+1}$. Therefore,
if $[{\rm dim}(G)+1]/h$ is not an integer, then the quantum and classical exponents $hM$ and
${{\rm dim}(G)+1}$ do not coincide. It would be interesting to understand how this information
is encoded in the dual description. Apparently, while on--shell we get the
analogue of \identityfive, there is no information about the relation \identityseven.
In particular, in the case of $Sp(N)$ the matrix model effective theory does not seem to
recognize the difference between the two critical exponents. We also note that whereas
in chiral correlators Eqs.\identityfive\ and  \identityseven\ may lead to the
same result, this is not the case of correlators which also contain non--chiral fields:
according to \identityseven\ these would be identically vanishing.

Let us conclude this note by observing that the above geometrical structure may help in
understanding the nature of the integrality condition on the periods of $T$. In this
respect, we note that integrality condition emerged in studying the eigenfunctions of
the Laplacian on a Riemann surface. In particular, in
\MatoneUY\ it has been shown that
eigenvalues with a nontrivial dependence on the complex structure
can be obtained as solutions of the equation
\eqn\ol{
\omega_{n',m'}=c\,\omega_{n,m}.} This
equation is equivalent to
\eqn\t{
m_j'-\sum_{k=1}^h\Omega_{jk}n_k'=\overline c
\left(m_j-\sum_{k=1}^h\Omega_{jk}n_k\right),\qquad j=1,\ldots,h,
}
 where $m_j,n_j,m_j',n_j'$ are
integers and $\Omega_{jk}$ is the Riemann period matrix.
In \MatoneUY\ it has been derived a set of solutions of such an
equation. The general problem involved in Eq.\t\
concerns the properties of the Riemann period matrix and its
number theoretical structure leading to periods with rational entries.
It can bee shown that such surfaces correspond to branched covering of the torus.
It would be interesting whether this is also the case of the Riemann surfaces
having integer period for $T$. In this respect we note
that holomorphic affine connections exist only on the torus. In the case of
branched covering of the torus this connection has poles but if this originated
{}from the holomorphic one of the base torus, this should reflect in peculiar
properties related to the integrality condition.

\vskip 20pt

\noindent {\bf Acknowledgements}. Work partially supported by the European
Community's Human Potential Programme under contract HPRN-CT-2000-00131 Quantum
Spacetime.

\listrefs

\end